\documentclass[aip,jcp,reprint,noshowkeys,superscriptaddress]{revtex4-1}
\usepackage{graphicx,dcolumn,bm,xcolor,microtype,multirow,amsmath,amssymb,amsfonts,physics,mhchem,xspace,subfigure}

\usepackage[T1]{fontenc}
\usepackage[utf8]{inputenc}

\usepackage{txfonts}

\usepackage[
	colorlinks=true,
    citecolor=blue,
    breaklinks=true
	]{hyperref}
\DeclareMathOperator*{\argmin}{arg\,min}

\newcommand{\ie}{\textit{i.e.}}

\definecolor{darkgreen}{HTML}{009900}
\usepackage[normalem]{ulem}

\newcommand{\SI}{\textcolor{blue}{supplementary material}}

\newcommand{\mc}{\multicolumn}

\newcommand{\tabc}[1]{\multicolumn{1}{c}{#1}}

\newcommand{\br}{\mathbf{r}}

\newcommand{\EPT}{E_{\text{PT2}}}
\newcommand{\EDMC}{E_{\text{FN-DMC}}}
\newcommand{\Ndet}{N_{\text{det}}}
\newcommand{\Nelec}{N}
\newcommand{\Nat}{M}
\newcommand{\hartree}{$E_h$}

\newcommand{\LCT}{Laboratoire de Chimie Th\'eorique (UMR 7616), Sorbonne Universit\'e, CNRS, Paris, France}
\newcommand{\ANL}{Computational Science Division, Argonne National Laboratory, Argonne, Illinois 60439, USA}
\newcommand{\LCPQ}{Laboratoire de Chimie et Physique Quantiques (UMR 5626), Universit\'e de Toulouse, CNRS, UPS, France}

\DeclareMathOperator{\erfc}{erfc}

\begin{document}

\title{Taming the fixed-node error in diffusion Monte Carlo via range separation}

\author{Anthony Scemama}
\email{scemama@irsamc.ups-tlse.fr}
\affiliation{\LCPQ}
\author{Emmanuel Giner}
\email{emmanuel.giner@lct.jussieu.fr}
\affiliation{\LCT}
\author{Anouar Benali}
\email{benali@anl.gov}
\affiliation{\ANL}
\author{Pierre-Fran\c{c}ois Loos}
\email{loos@irsamc.ups-tlse.fr}
\affiliation{\LCPQ}

\begin{abstract}
By combining density-functional theory (DFT) and wave function theory (WFT) via the range separation (RS) of the interelectronic Coulomb operator, we obtain accurate fixed-node diffusion Monte Carlo (FN-DMC) energies with compact multi-determinant trial wave functions.
In particular, we combine here short-range exchange-correlation functionals with a flavor of selected configuration interaction (SCI) known as \emph{configuration interaction using a perturbative selection made iteratively} (CIPSI), a scheme that we label RS-DFT-CIPSI.
One of the take-home messages of the present study is that RS-DFT-CIPSI trial wave functions yield lower fixed-node energies with more compact multi-determinant expansions than CIPSI, especially for small basis sets.
Indeed, as the CIPSI component of RS-DFT-CIPSI is relieved from describing the short-range part of the correlation hole around the electron-electron coalescence points, the number of determinants in the trial wave function required to reach a given accuracy is significantly reduced as compared to a conventional CIPSI calculation.
Importantly, by performing various numerical experiments, we evidence that the RS-DFT scheme essentially plays the role of a simple Jastrow factor by mimicking short-range correlation effects, hence avoiding the burden of performing a stochastic optimization.
Considering the 55 atomization energies of the Gaussian-1 benchmark set of molecules, we show that using a fixed value of $\mu=0.5$~bohr$^{-1}$ provides effective error cancellations as well as compact trial wave functions, making the present method a good candidate for the accurate description of large chemical systems.
\end{abstract}

\maketitle

\section{Introduction}
\label{sec:intro}

Solving the Schr\"odinger equation for the ground state of atoms and molecules is a complex task that has kept theoretical and computational chemists busy for almost a hundred years now. \cite{Schrodinger_1926}
In order to achieve this formidable endeavor, various strategies have been carefully designed and efficiently implemented in various quantum chemistry software packages.

\subsection{Wave function-based methods}

One of these strategies consists in relying on wave function theory \cite{Pople_1999} (WFT) and, in particular, on the full configuration interaction (FCI) method.
However, FCI delivers only the exact solution of the Schr\"odinger equation within a finite basis (FB) of one-electron functions, the FB-FCI energy being an upper bound to the exact energy in accordance with the variational principle.
The FB-FCI wave function and its corresponding energy form the eigenpair of an approximate Hamiltonian defined as
the projection of the exact Hamiltonian onto the finite many-electron basis of
all possible Slater determinants generated within this finite one-electron basis.
The FB-FCI wave function can then be interpreted as a constrained solution of the
true Hamiltonian forced to span the restricted space provided by the finite one-electron basis.
In the complete basis set (CBS) limit, the constraint is lifted and the exact energy and wave function are recovered.
Hence, the accuracy of a FB-FCI calculation can be systematically improved by increasing the size of the one-electron basis set.
Nevertheless, the exponential growth of its computational cost with the number of electrons and with the basis set size is prohibitive for most chemical systems.

In recent years, the introduction of new algorithms \cite{White_1992,Booth_2009,Thom_2010,Xu_2018,Motta_2018,Deustua_2018,Eriksen_2018,Eriksen_2019,Ghanem_2019} and the
revival \cite{Abrams_2005,Bytautas_2009,Roth_2009,Giner_2013,Knowles_2015,Holmes_2016,Holmes_2017,Sharma_2017,Evangelista_2014,Liu_2016,Tubman_2016,Tubman_2020,Per_2017,Zimmerman_2017,Ohtsuka_2017,Garniron_2018}
of selected configuration interaction (SCI)
methods \cite{Bender_1969,Huron_1973,Buenker_1974} significantly expanded the range of applicability of this family of methods.
Importantly, one can now routinely compute the ground- and excited-state energies of small- and medium-sized molecular systems with near-FCI accuracy. \cite{Booth_2010,Cleland_2010,Daday_2012,Motta_2017,Chien_2018,Loos_2018a,Loos_2019,Loos_2020b,Loos_2020c,Williams_2020,Eriksen_2020}
However, although the prefactor is reduced, the overall computational scaling remains exponential unless some bias is introduced leading
to a loss of size consistency. \cite{Evangelisti_1983,Cleland_2010,Tenno_2017,Ghanem_2019}

\subsection{Density-based methods}

Another route to solve the Schr\"odinger equation is density-functional theory (DFT). \cite{Hohenberg_1964,Kohn_1999} 
Present-day DFT calculations are almost exclusively done within the so-called Kohn-Sham (KS) formalism, \cite{Kohn_1965} which
transfers the complexity of the many-body problem to the universal and yet unknown exchange-correlation (xc) functional thanks to a judicious mapping between a non-interacting reference system and its interacting analog which both have the same one-electron density.
KS-DFT \cite{Hohenberg_1964,Kohn_1965} is now the workhorse of electronic structure calculations for atoms, molecules and solids thanks to its very favorable accuracy/cost ratio. \cite{ParrBook}
As compared to WFT, DFT has the indisputable advantage of converging much faster with respect to the size of the basis set. \cite{FraMusLupTou-JCP-15,Giner_2018,Loos_2019d,Giner_2020}
However, unlike WFT where, for example, many-body perturbation theory provides a precious tool to go toward the exact wave function, there is no systematic way to improve approximate xc functionals toward the exact functional.
Therefore, one faces, in practice, the unsettling choice of the \emph{approximate} xc functional. \cite{Becke_2014}
Moreover, because of the approximate nature of the xc functional, although the resolution of the KS equations is variational, the resulting KS energy does not have such property.

\subsection{Stochastic methods}

Diffusion Monte Carlo (DMC) belongs to the family of stochastic methods known as quantum Monte Carlo (QMC) and is yet another numerical scheme to obtain
the exact solution of the Schr\"odinger equation with a different
twist. \cite{Foulkes_2001,Austin_2012,Needs_2020} 
In DMC, the solution is imposed to have the same nodes (or zeroes)
as a given (approximate) antisymmetric trial wave function. \cite{Reynolds_1982,Ceperley_1991}
Within this so-called fixed-node (FN) approximation,
the FN-DMC energy associated with a given trial wave function is an upper
bound to the exact energy, and the latter is recovered only when the
nodes of the trial wave function coincide with the nodes of the exact
wave function.
The trial wave function, which can be single- or multi-determinantal in nature depending on the type of correlation at play and the target accuracy, is the key ingredient dictating, via the quality of its nodal surface, the accuracy of the resulting energy and properties.

The polynomial scaling of its computational cost with respect to the number of electrons and with the size
of the trial wave function makes the FN-DMC method particularly attractive.
This favorable scaling, its very low memory requirements and
its adequacy with massively parallel architectures make it a
serious alternative for high-accuracy simulations of large systems. \cite{Nakano_2020,Scemama_2013,Needs_2020,Kim_2018,Kent_2020}
In addition, the total energies obtained are usually far below
those obtained with the FCI method in computationally tractable basis
sets because the constraints imposed by the fixed-node approximation
are less severe than the constraints imposed by the finite-basis
approximation.
However, because it is not possible to minimize directly the FN-DMC energy with respect
to the linear and non-linear parameters of the trial wave function, the
fixed-node approximation is much more difficult to control than the
finite-basis approximation, especially to compute energy differences.
The conventional approach consists in multiplying the determinantal part of the trial wave
function by a positive function, the Jastrow factor, which main assignment is to take into
account the bulk of the dynamical electron correlation and reduce the statistical fluctuations without altering the location of the nodes.
The determinantal part of the trial wave function is then stochastically re-optimized within variational
Monte Carlo (VMC) in the presence of the Jastrow factor (which can also be simultaneously optimized) and the nodal
surface is expected to be improved. \cite{Umrigar_2005,Scemama_2006,Umrigar_2007,Toulouse_2007,Toulouse_2008}
Using this technique, it has been shown that the chemical accuracy could be reached within
FN-DMC.\cite{Petruzielo_2012}

\subsection{Single-determinant trial wave functions}
\label{sec:SD}

The qualitative picture of the electronic structure of weakly
correlated systems, such as organic molecules near their equilibrium
geometry, is usually well represented with a single Slater
determinant. This feature is in part responsible for the success of
DFT and coupled cluster (CC) theory.
Likewise, DMC with a single-determinant trial wave function can be used as a
single-reference post-Hartree-Fock method for weakly correlated systems, with an accuracy comparable
to CCSD(T), \cite{Dubecky_2014,Grossman_2002} the gold standard of WFT for ground state energies. \cite{Cizek_1969,Purvis_1982}
In single-determinant DMC calculations, the only degree of freedom available to
reduce the fixed-node error are the molecular orbitals with which the
Slater determinant is built.
Different molecular orbitals can be chosen:
Hartree-Fock (HF), Kohn-Sham (KS), natural orbitals (NOs) of a
correlated wave function, or orbitals optimized in the
presence of a Jastrow factor.
Nodal surfaces obtained with a KS determinant are in general
better than those obtained with a HF determinant,\cite{Per_2012} and
of comparable quality to those obtained with a Slater determinant
built with NOs.\cite{Wang_2019} Orbitals obtained in the presence
of a Jastrow factor are generally superior to KS
orbitals.\cite{Filippi_2000,Scemama_2006,HaghighiMood_2017,Ludovicy_2019}

The description of electron correlation within DFT is very different
from correlated methods such as FCI or CC.
As mentioned above, within KS-DFT, one solves a mean-field problem 
with a modified potential incorporating the effects of electron correlation 
while maintaining the exact ground state density, whereas in
correlated methods the real Hamiltonian is used and the
electron-electron interaction is explicitly considered.
Nevertheless, as the orbitals are one-electron functions,
the procedure of orbital optimization in the presence of a
Jastrow factor can be interpreted as a self-consistent field procedure
with an effective Hamiltonian,\cite{Filippi_2000} similarly to DFT.
So KS-DFT can be viewed as a very cheap way of introducing the effect of
correlation in the orbital coefficients dictating the location of the nodes of a single Slater determinant.
Yet, even when employing the exact xc potential in a complete basis set, a fixed-node error necessarily remains because the
single-determinant ans\"atz does not have enough flexibility for describing the
nodal surface of the exact correlated wave function for a generic many-electron
system. \cite{Ceperley_1991,Bressanini_2012,Loos_2015b}
If one wants to recover the exact energy, a multi-determinant parameterization
of the wave functions must be considered.

\subsection{Multi-determinant trial wave functions}

The single-determinant trial wave function approach obviously fails in the presence of strong correlation, like in 
transition metal complexes, low-spin open-shell systems, and covalent bond breaking situations which cannot be qualitatively described by a single electronic configuration.
In such cases or when very high accuracy is required, a viable alternative is to consider the FN-DMC method as a
``post-FCI'' method. A multi-determinant trial wave function is then produced by
approaching FCI with a SCI method such as \emph{configuration interaction using a perturbative
selection made iteratively} (CIPSI). \cite{Giner_2013,Giner_2015,Caffarel_2016_2}
When the basis set is enlarged, the trial wave function gets closer to
the exact wave function, so we expect the nodal surface to be
improved.\cite{Caffarel_2016}
Note that, as discussed in Ref.~\onlinecite{Caffarel_2016_2}, there is no mathematical guarantee that increasing the size of the one-electron basis lowers the FN-DMC energy, because the variational principle does not explicitly optimize the nodal surface, nor the FN-DMC energy.
However, in all applications performed so far, \cite{Giner_2013,Scemama_2014,Scemama_2016,Giner_2015,Caffarel_2016,Scemama_2018,Scemama_2018b,Scemama_2019} a systematic decrease of the FN-DMC energy has been observed whenever the SCI trial wave function is improved variationally upon enlargement of the basis set.

The technique relying on CIPSI multi-determinant trial wave functions described above has the advantage of using near-FCI quality nodes in a given basis
set, which is perfectly well defined and therefore makes the calculations systematically improvable and reproducible in a
black-box way without needing any QMC expertise.
Nevertheless, this procedure cannot be applied to large systems because of the
exponential growth of the number of Slater determinants in the trial wave function.
Extrapolation techniques have been employed to estimate the FN-DMC energies
obtained with FCI wave functions,\cite{Scemama_2018,Scemama_2018b,Scemama_2019} and other authors
have used a combination of the two approaches where highly truncated
CIPSI trial wave functions are stochastically re-optimized in VMC under the presence
of a Jastrow factor to keep the number of determinants
small,\cite{Giner_2016} and where the consistency between the
different wave functions is kept by imposing a constant energy
difference between the estimated FCI energy and the variational energy
of the SCI wave function.\cite{Dash_2018,Dash_2019}
Nevertheless, finding a robust protocol to obtain high accuracy
calculations which can be reproduced systematically and
applicable to large systems with a multi-configurational character is
still an active field of research. The present paper falls
within this context.

The central idea of the present work, and the launch pad for the remainder of this study, is that one can combine the various strengths of WFT, DFT, and QMC in order to create a new hybrid method with more attractive features and higher accuracy.
In particular, we show here that one can combine CIPSI and KS-DFT via the range separation (RS) of the interelectronic Coulomb operator \cite{Sav-INC-96a,Toulouse_2004} --- a scheme that we label RS-DFT-CIPSI in the following --- to obtain accurate FN-DMC energies with compact multi-determinant trial wave functions.
An important take-home message from the present study is that the RS-DFT scheme essentially plays the role of a simple Jastrow factor by mimicking short-range correlation effects.
Thanks to this, RS-DFT-CIPSI multi-determinant trial wave functions yield lower fixed-node energies with more compact multi-determinant expansion than CIPSI, especially for small basis sets, and can be produced in a completely deterministic and systematic way, without the burden of the stochastic optimization.

The present manuscript is organized as follows.
In Sec.~\ref{sec:rsdft-cipsi}, we provide theoretical details about the CIPSI algorithm (Sec.~\ref{sec:CIPSI}) and range-separated DFT (Sec.~\ref{sec:rsdft}).
Computational details are reported in Sec.~\ref{sec:comp-details}.
In Sec.~\ref{sec:mu-dmc}, we discuss the influence of the range-separation parameter on the fixed-node error as well as the link between RS-DFT and Jastrow factors.
Section \ref{sec:atomization} examines the performance of the present scheme for the atomization energies of the Gaussian-1 set of molecules.
Finally, we draw our conclusion in Sec.~\ref{sec:conclusion}.
Unless otherwise stated, atomic units are used.

\section{Theory}
\label{sec:rsdft-cipsi}

\subsection{The CIPSI algorithm}
\label{sec:CIPSI}
Beyond the single-determinant representation, the best
multi-determinant wave function one can wish for --- in a given basis set --- is the FCI wave function.
FCI is the ultimate goal of post-HF methods, and there exist several systematic
improvements on the path from HF to FCI:
i) increasing the maximum degree of excitation of CI methods (CISD, CISDT,
CISDTQ,~\ldots), or ii) expanding the size of a complete active space
(CAS) wave function until all the orbitals are in the active space.
SCI methods take a shortcut between the HF
determinant and the FCI wave function by increasing iteratively the
number of determinants on which the wave function is expanded,
selecting the determinants which are expected to contribute the most
to the FCI wave function. At each iteration, the lowest eigenpair is
extracted from the CI matrix expressed in the determinant subspace,
and the FCI energy can be estimated by adding up to the variational energy 
a second-order perturbative correction (PT2), $\EPT$.
The magnitude of $\EPT$ is a measure of the distance to the FCI energy 
and a diagnostic of the quality of the wave function.
Within the CIPSI algorithm originally developed by Huron \textit{et al.} 
in Ref.~\onlinecite{Huron_1973} and efficiently implemented in \emph{Quantum
Package} as described in Ref.~\onlinecite{Garniron_2019}, the PT2
correction is computed simultaneously to the determinant selection at no extra cost. 
$\EPT$ is then the sole parameter of the CIPSI algorithm and is chosen to be its convergence criterion.

\subsection{Range-separated DFT}
\label{sec:rsdft}

Range-separated DFT (RS-DFT) was introduced in the seminal work of Savin. \cite{Sav-INC-96a,Toulouse_2004}
In RS-DFT, the Coulomb operator entering the electron-electron repulsion is split into two pieces:
\begin{equation}
  \frac{1}{r}
  = w_{\text{ee}}^{\text{sr}, \mu}(r)
  + w_{\text{ee}}^{\text{lr}, \mu}(r),
\end{equation}
where
\begin{align}
  w_{\text{ee}}^{\text{sr},\mu}(r) & = \frac{\erfc \qty( \mu\, r)}{r},
  &
  w_{\text{ee}}^{\text{lr},\mu}(r) & = \frac{\erf \qty( \mu\, r)}{r}
\end{align}
are the singular short-range (sr) part and the non-singular long-range (lr) part, respectively, $\mu$ is the range-separation parameter which controls how rapidly the short-range part decays, $\erf(x)$ is the error function, and $\erfc(x) = 1 - \erf(x)$ is its complementary version.

The main idea behind RS-DFT is to treat the short-range part of the
interaction using a density functional, and the long-range part within a WFT method like FCI in the present case.
The parameter $\mu$ controls the range of the separation, and allows
to go continuously from the KS Hamiltonian ($\mu=0$) to
the FCI Hamiltonian ($\mu = \infty$).

To rigorously connect WFT and DFT, the universal
Levy-Lieb density functional \cite{Lev-PNAS-79,Lie-IJQC-83} is
decomposed as
\begin{equation}
  \mathcal{F}[n] = \mathcal{F}^{\text{lr},\mu}[n] + \bar{E}_{\text{Hxc}}^{\text{sr,}\mu}[n],
  \label{Fdecomp}
\end{equation}
where $n$ is a one-electron density,
$\mathcal{F}^{\text{lr},\mu}$ is a long-range universal density
functional and $\bar{E}_{\text{Hxc}}^{\text{sr,}\mu}$ is the
complementary short-range Hartree-exchange-correlation (Hxc) density
functional. \cite{Savin_1996,Toulouse_2004}
The exact ground state energy can be therefore obtained as a minimization
over a multi-determinant wave function as follows:
\begin{equation}
  \label{min_rsdft} E_0= \min_{\Psi} \qty{
  \mel{\Psi}{\hat{T}+\hat{W}_\text{{ee}}^{\text{lr},\mu}+\hat{V}_{\text{ne}}}{\Psi}
  + \bar{E}^{\text{sr},\mu}_{\text{Hxc}}[n_\Psi]
  },
\end{equation}
with $\hat{T}$ the kinetic energy operator,
$\hat{W}_\text{ee}^{\text{lr},\mu}$ the long-range
electron-electron interaction,
$n_\Psi$ the one-electron density associated with $\Psi$,
and $\hat{V}_{\text{ne}}$ the electron-nucleus potential.
The minimizing multi-determinant wave function $\Psi^\mu$
can be determined by the self-consistent eigenvalue equation
\begin{equation}
  \label{rs-dft-eigen-equation}
  \hat{H}^\mu[n_{\Psi^{\mu}}] \ket{\Psi^{\mu}}= \mathcal{E}^{\mu} \ket{\Psi^{\mu}},
\end{equation}
with the long-range interacting Hamiltonian
\begin{equation}
  \label{H_mu}
  \hat{H}^\mu[n_{\Psi^{\mu}}] = \hat{T}+\hat{W}_{\text{ee}}^{\text{lr},\mu}+\hat{V}_{\text{ne}}+ \hat{\bar{V}}_{\text{Hxc}}^{\text{sr},\mu}[n_{\Psi^{\mu}}],
\end{equation}
where
$\hat{\bar{V}}_{\text{Hxc}}^{\text{sr},\mu}$
is the complementary short-range Hartree-exchange-correlation
potential operator.
Once $\Psi^{\mu}$ has been calculated, the electronic ground-state
energy is obtained as
\begin{equation}
  \label{E-rsdft}
  E_0=  \mel{\Psi^{\mu}}{\hat{T}+\hat{W}_\text{{ee}}^{\text{lr},\mu}+\hat{V}_{\text{ne}}}{\Psi^{\mu}}+\bar{E}^{\text{sr},\mu}_{\text{Hxc}}[n_{\Psi^\mu}].
\end{equation}

Note that, for $\mu=0$, the long-range interaction vanishes, \ie,
$w_{\text{ee}}^{\text{lr},\mu=0}(r) = 0$, and thus RS-DFT reduces to standard
KS-DFT and $\Psi^\mu$ is the KS determinant. For $\mu = \infty$, the long-range
interaction becomes the standard Coulomb interaction, \ie,
$w_{\text{ee}}^{\text{lr},\mu\to\infty}(r) = r^{-1}$, and thus RS-DFT reduces
to standard WFT and $\Psi^\mu$ is the FCI wave function.

\begin{figure*}
  \includegraphics[width=0.7\linewidth]{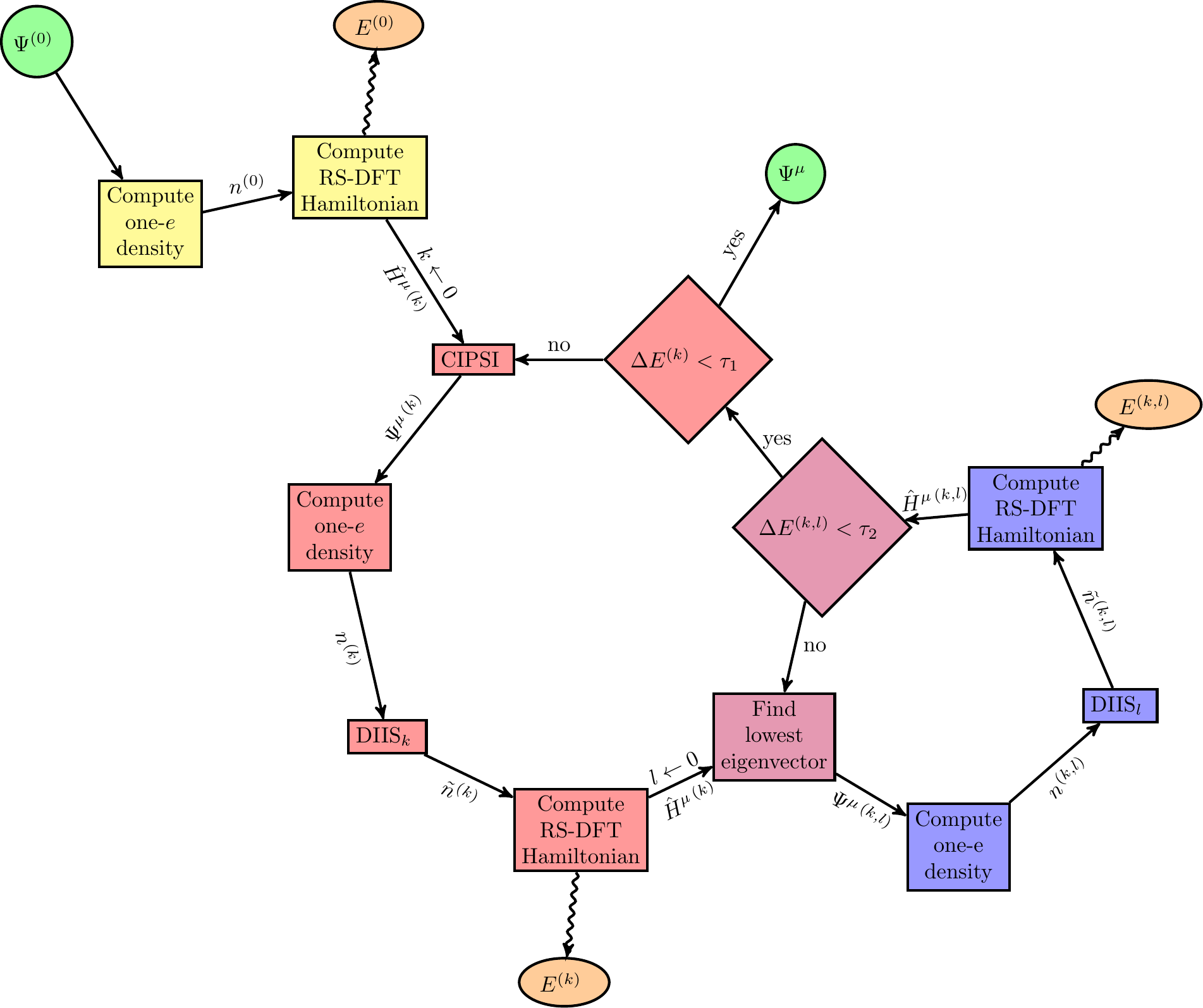}
  \caption{Algorithm showing the generation of the RS-DFT wave
    function $\Psi^{\mu}$ starting from $\Psi^{(0)}$. 
    The outer (macro-iteration) and inner (micro-iteration) loops are represented in red and blue, respectively.
    The steps common to both loops are represented in purple.
    DIIS extrapolations of the one-electron density are introduced in both the outer and inner loops in order to speed up convergence of the self-consistent process.}
  \label{fig:algo}
\end{figure*}

Hence, range separation creates a continuous path connecting smoothly the KS determinant to the
FCI wave function. Because the KS nodes are of higher quality than the
HF nodes (see Sec.~\ref{sec:SD}), we expect that using wave functions built along this path
will always provide reduced fixed-node errors compared to the path
connecting HF to FCI which consists in increasing the number of determinants.

We follow the KS-to-FCI path by performing FCI calculations using the
RS-DFT Hamiltonian with different values of $\mu$.
Our algorithm, depicted in Fig.~\ref{fig:algo}, starts with a
single- or multi-determinant wave function $\Psi^{(0)}$ which can 
be obtained in many different ways depending on the system that one considers.
One of the particularity of the present work is that we
use the CIPSI algorithm to perform approximate FCI calculations
with the RS-DFT Hamiltonian $\hat{H}^\mu$. \cite{Giner_2018}
This provides a multi-determinant trial wave function $\Psi^{\mu}$ that one can ``feed'' to DMC.
In the outer (macro-iteration) loop (red), at the $k$th iteration, a CIPSI selection is performed 
to obtain $\Psi^{\mu\,(k)}$ with the RS-DFT Hamiltonian $\hat{H}^{\mu\,(k)}$
parameterized using the current one-electron density $n^{(k)}$.
At each iteration, the number of determinants in $\Psi^{\mu\,(k)}$ increases.
One exits the outer loop when the absolute energy difference between two successive macro-iterations $\Delta E^{(k)}$ is below a threshold $\tau_1$ that has been set to $10^{-3}$ \hartree{} in the present study and which is consistent with the CIPSI threshold (see Sec.~\ref{sec:comp-details}).
An inner (micro-iteration) loop (blue) is introduced to accelerate the
convergence of the self-consistent calculation, in which the set of
determinants in $\Psi^{\mu\,(k,l)}$ is kept fixed, and only the diagonalization of
$\hat{H}^{\mu\,(k,l)}$ is performed iteratively with the updated density $n^{(k,l)}$.
The inner loop is exited when the absolute energy difference between two successive micro-iterations $\Delta E^{(k,l)}$ is below a threshold $\tau_2$ that has been here set to $10^{-2} \times \tau_1$.
The convergence of the algorithm was further improved
by introducing a direct inversion in the iterative subspace (DIIS)
step to extrapolate the one-electron density both in the outer and inner loops. \cite{Pulay_1980,Pulay_1982}
We emphasize that any range-separated post-HF method can be
implemented using this scheme by just replacing the CIPSI step by the
post-HF method of interest.
Note that, thanks to the self-consistent nature of the algorithm,
the final trial wave function $\Psi^{\mu}$ is independent of the starting wave function $\Psi^{(0)}$.

\section{Computational details}
\label{sec:comp-details}

All reference data (geometries, atomization
energies, zero-point energy, etc) were taken from the NIST
computational chemistry comparison and benchmark database
(CCCBDB).\cite{nist}
In the reference atomization energies, the zero-point vibrational
energy was removed from the experimental atomization energies.

All calculations have been performed using Burkatzki-Filippi-Dolg (BFD)
pseudopotentials \cite{Burkatzki_2007,Burkatzki_2008} with the associated double-,
triple-, and quadruple-$\zeta$ basis sets (V$X$Z-BFD).
The small-core BFD pseudopotentials include scalar relativistic effects.
Coupled cluster with singles, doubles, and perturbative triples [CCSD(T)] \cite{Scuseria_1988,Scuseria_1989} and KS-DFT energies have been computed with
\emph{Gaussian09},\cite{g16} using the unrestricted formalism for open-shell systems.

The CIPSI calculations have been performed with \emph{Quantum
Package}.\cite{Garniron_2019,qp2_2020} We consider the short-range version
of the local-density approximation (LDA)\cite{Sav-INC-96a,TouSavFla-IJQC-04} and Perdew-Burke-Ernzerhof (PBE) \cite{PerBurErn-PRL-96}
xc functionals defined in
Ref.~\onlinecite{GolWerStoLeiGorSav-CP-06} (see also
Refs.~\onlinecite{TouColSav-JCP-05,GolWerSto-PCCP-05}) that we label srLDA and srPBE respectively in the following.
In this work, we target chemical accuracy, so 
the convergence criterion for stopping the CIPSI calculations
has been set to $\EPT < 10^{-3}$ \hartree{} or $ \Ndet > 10^7$.
All the wave functions are eigenfunctions of the $\Hat{S}^2$ spin operator, as
described in Ref.~\onlinecite{Applencourt_2018}.

QMC calculations have been performed with \textit{QMC=Chem},\cite{Scemama_2013}
in the determinant localization approximation (DLA),\cite{Zen_2019}
where only the determinantal component of the trial wave
function is present in the expression of the wave function on which
the pseudopotential is localized. Hence, in the DLA, the fixed-node
energy is independent of the Jastrow factor, as in all-electron
calculations. Simple Jastrow factors were used to reduce the
fluctuations of the local energy (see Sec.~\ref{sec:rsdft-j} for their explicit expression).
The FN-DMC simulations are performed with all-electron moves using the
stochastic reconfiguration algorithm developed by Assaraf \textit{et al.}
\cite{Assaraf_2000} with a time step of $5 \times 10^{-4}$ a.u.,
independent populations of 100 walkers and a projecting time of $1$
a.u. With such parameters, both the time-step error and the
bias due to the finite projecting time are
smaller than the error bars.

All the data related to the present study (geometries, basis sets, total energies, \textit{etc}) can be found in the {\SI}.

\section{Influence of the range-separation parameter on the fixed-node error}
\label{sec:mu-dmc}

\begin{table}
  \caption{FN-DMC energy $\EDMC$ (in \hartree{}) and number of determinants $\Ndet$ in \ce{H2O} for various trial wave functions $\Psi^{\mu}$ obtained with the srPBE density functional.}
  \label{tab:h2o-dmc}
  \begin{ruledtabular}
  \begin{tabular}{crlrl}
	&         \multicolumn{2}{c}{VDZ-BFD}  &           \multicolumn{2}{c}{VTZ-BFD}  \\
		\cline{2-3} \cline{4-5}
	$\mu$     &  $\Ndet$     & $\EDMC$           &  $\Ndet$        &  $\EDMC$           \\
\hline
  $0.00$    &  $11$        & $-17.253\,59(6)$  &  $23$           &  $-17.256\,74(7)$  \\
  $0.20$    &  $23$        & $-17.253\,73(7)$  &  $23$           &  $-17.256\,73(8)$  \\
  $0.30$    &  $53$        & $-17.253\,4(2)$   &  $219$          &  $-17.253\,7(5)$   \\
  $0.50$    &  $1\,442$    & $-17.253\,9(2)$   &  $16\,99$       &  $-17.257\,7(2)$   \\
  $0.75$    &  $3\,213$    & $-17.255\,1(2)$   &  $13\,362$      &  $-17.258\,4(3)$   \\
  $1.00$    &  $6\,743$    & $-17.256\,6(2)$   &  $256\,73$      &  $-17.261\,0(2)$   \\
  $1.75$    &  $54\,540$   & $-17.259\,5(3)$   &  $207\,475$     &  $-17.263\,5(2)$   \\
  $2.50$    &  $51\,691$   & $-17.259\,4(3)$   &  $858\,123$     &  $-17.264\,3(3)$   \\
  $3.80$    &  $103\,059$  & $-17.258\,7(3)$   &  $1\,621\,513$  &  $-17.263\,7(3)$   \\
  $5.70$    &  $102\,599$  & $-17.257\,7(3)$   &  $1\,629\,655$  &  $-17.263\,2(3)$   \\
  $8.50$    &  $101\,803$  & $-17.257\,3(3)$   &  $1\,643\,301$  &  $-17.263\,3(4)$   \\
  $\infty$  &  $200\,521$  & $-17.256\,8(6)$   &  $1\,631\,982$  &  $-17.263\,9(3)$   \\
  \end{tabular}
  \end{ruledtabular}
\end{table}

\begin{figure}
  \includegraphics[width=\columnwidth]{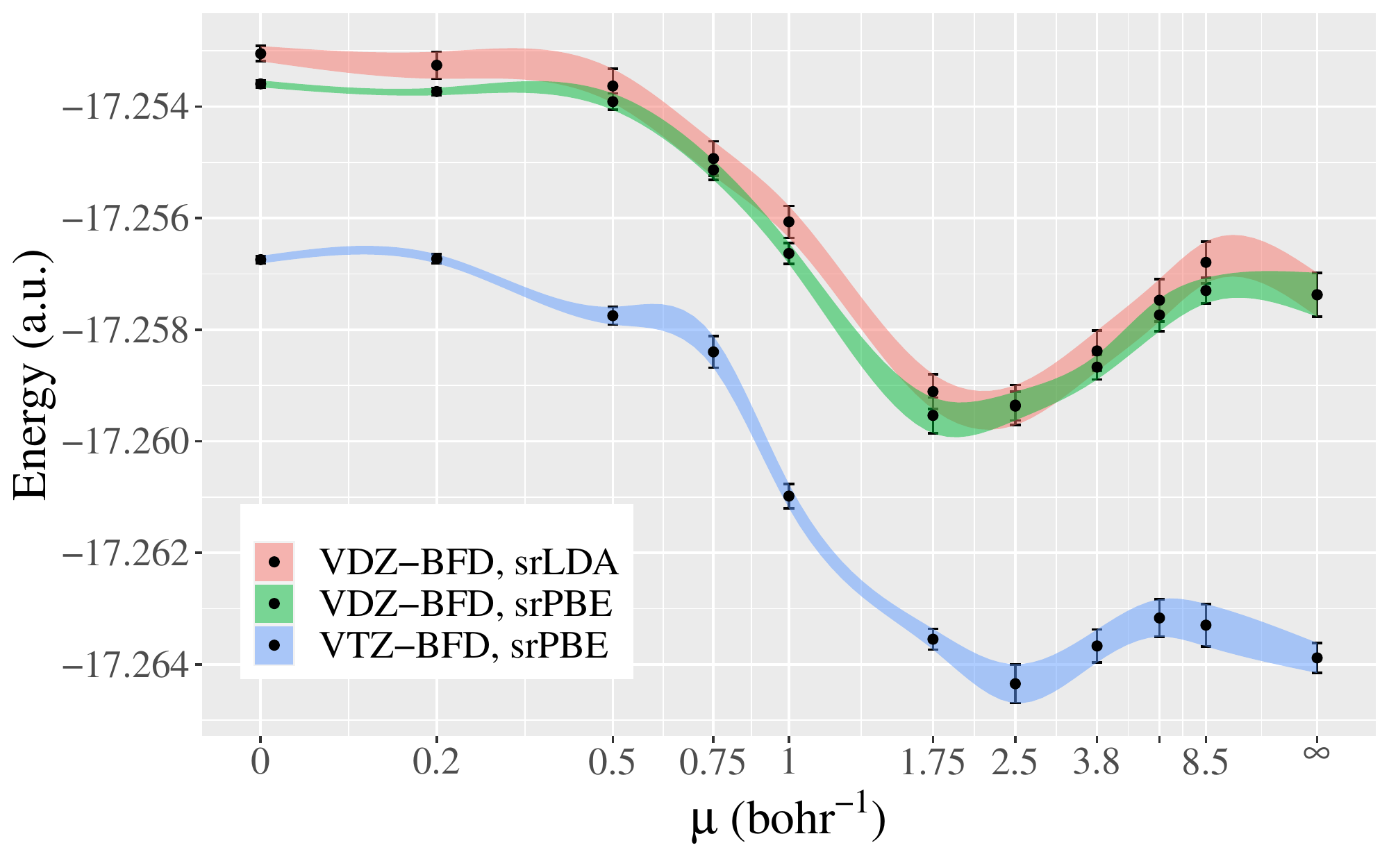}
  \caption{FN-DMC energy of \ce{H2O} as a function 
    of $\mu$ for various trial wave functions $\Psi^{\mu}$ generated at different levels of theory.
    The raw data can be found in the {\SI}.}
  \label{fig:h2o-dmc}
\end{figure}

The first question we would like to address is the quality of the
nodes of the wave function $\Psi^{\mu}$ obtained for intermediate values of the
range separation parameter (\ie, $0 < \mu  < +\infty$).
For this purpose, we consider a weakly correlated molecular system, namely the water
molecule at its experimental geometry. \cite{Caffarel_2016}
We then generate trial wave functions $\Psi^\mu$ for multiple values of
$\mu$, and compute the associated FN-DMC energy keeping fixed all the
parameters impacting the nodal surface, such as the CI coefficients and the molecular orbitals.

\subsection{Fixed-node energy of RS-DFT-CIPSI trial wave functions}
\label{sec:fndmc_mu}
From Table~\ref{tab:h2o-dmc} and Fig.~\ref{fig:h2o-dmc}, where we report the fixed-node energy of \ce{H2O} as a function of $\mu$ for various short-range density functionals and basis sets,
one can clearly observe that relying on FCI trial
wave functions ($\mu = \infty$) give FN-DMC energies lower
than the energies obtained with a single KS determinant ($\mu=0$):
a lowering of $3.2 \pm 0.6$~m\hartree{} at the double-$\zeta$ level and $7.2 \pm
0.3$~m\hartree{} at the triple-$\zeta$ level are obtained with the srPBE functional.
Coming now to the nodes of the trial wave function $\Psi^{\mu}$ with
intermediate values of $\mu$, Fig.~\ref{fig:h2o-dmc} shows that
a smooth behavior is obtained:
starting from $\mu=0$ (\ie, the KS determinant),
the FN-DMC error is reduced continuously until it reaches a minimum 
for an optimal value of $\mu$ (which is obviously basis set and functional dependent),
and then the FN-DMC error raises until it reaches the $\mu=\infty$ limit (\ie, the FCI wave function).
For instance, with respect to the fixed-node energy associated with the RS-DFT-CIPSI(srPBE/VDZ-BFD) trial wave function at $\mu=\infty$, 
one can obtain a lowering of the FN-DMC energy of $2.6 \pm 0.7$~m\hartree{}
with an optimal value of $\mu=1.75$~bohr$^{-1}$. 
This lowering in FN-DMC energy is to be compared with the $3.2 \pm
0.7$~m\hartree{} gain in FN-DMC energy between the KS wave function ($\mu=0$)
and the FCI wave function ($\mu=\infty$).  When the basis set is improved, the
gain in FN-DMC energy with respect to the FCI trial wave function is reduced,
and the optimal value of $\mu$ is slightly shifted towards large $\mu$ as expected. 
Last but not least, the nodes of the wave functions $\Psi^\mu$ obtained with the srLDA 
functional give very similar FN-DMC energies with respect
to those obtained with srPBE, even if the
RS-DFT energies obtained with these two functionals differ by several
tens of m\hartree{}. 
Accordingly, all the RS-DFT calculations are performed with the srPBE functional in the remaining of this paper.

Another important aspect here is the compactness of the trial wave functions $\Psi^\mu$:
at $\mu=1.75$~bohr$^{-1}$, $\Psi^{\mu}$ has \textit{only} $54\,540$ determinants at the RS-DFT-CIPSI(srPBE/VDZ-BFD) level, while the FCI wave function contains $200\,521$ determinants (see Table \ref{tab:h2o-dmc}). Even at the RS-DFT-CIPSI(srPBE/VTZ-BFD) level, we observe a reduction by a factor two in the number of determinants between the optimal $\mu$ value and $\mu = \infty$.
The take-home message of this first numerical study is that RS-DFT-CIPSI trial wave functions can yield a lower fixed-node energy with more compact multi-determinant expansion as compared to FCI.
This is a key result of the present study.

\subsection{RS-DFT vs Jastrow factor}
\label{sec:rsdft-j}
The data presented in Sec.~\ref{sec:fndmc_mu} evidence that, in a finite basis, RS-DFT can provide 
trial wave functions with better nodes than FCI wave functions.
As mentioned in Sec.~\ref{sec:SD}, such behavior can be directly compared to the common practice of
re-optimizing the multi-determinant part of a trial wave function $\Psi$ (the so-called Slater part) in the presence of the exponentiated Jastrow factor $e^J$. \cite{Umrigar_2005,Scemama_2006,Umrigar_2007,Toulouse_2007,Toulouse_2008}
Hence, in the present paragraph, we would like to elaborate further on the link between RS-DFT
and wave function optimization in the presence of a Jastrow factor.
For the sake of simplicity, the molecular orbitals and the Jastrow
factor are kept fixed; only the CI coefficients are varied.

Let us then assume a fixed Jastrow factor $J(\br_1, \ldots , \br_\Nelec)$ (where $\br_i$ is the position of the $i$th electron and $\Nelec$ the total number of electrons),
and a corresponding Slater-Jastrow wave function $\Phi = e^J \Psi$,
where 
\begin{equation}
\label{eq:Slater}
  \Psi = \sum_I c_I D_I
\end{equation} 
is a general linear combination of (fixed) Slater determinants $D_I$.
The only variational parameters in $\Phi$ are therefore the coefficients $c_I$ belonging to the Slater part $\Psi$.
Let us define $\Psi^J$ as the linear combination of Slater determinants minimizing the variational energy associated with $\Phi$, \ie,
\begin{equation}
 \Psi^J = \argmin_{\Psi}\frac{ \mel{ \Psi }{ e^{J} \hat{H} e^{J} }{ \Psi } }{\mel{ \Psi }{ e^{2J} }{ \Psi } }.
\end{equation}
Such a wave function satisfies the generalized Hermitian eigenvalue equation
\begin{equation}
 e^{J} \hat{H} \qty( e^{J} \Psi^J ) = E \, e^{2J} \Psi^J,
\label{eq:ci-j}
\end{equation}
but also the non-Hermitian transcorrelated eigenvalue problem\cite{BoyHan-PRSLA-69,BoyHanLin-1-PRSLA-69,BoyHanLin-2-PRSLA-69,Tenno_2000,Luo-JCP-10,YanShi-JCP-12,CohLuoGutDowTewAla-JCP-19}
\begin{equation}
 \label{eq:transcor}
 e^{-J}  \hat{H} \qty( e^{J} \Psi^J) = E \, \Psi^J,
\end{equation}
which is much easier to handle despite its non-Hermiticity.
Of course, the FN-DMC energy of $\Phi$ depends only on the nodes of $\Psi^J$ as the positivity of the Jastrow factor makes sure that it does not alter the nodal surface.
In a finite basis set and with an accurate Jastrow factor, it is known that the nodes
of $\Psi^J$ may be better than the nodes of the FCI wave function. 
Hence, we would like to compare $\Psi^J$ and $\Psi^\mu$.

To do so, we have made the following numerical experiment.
First, we extract the 200 determinants with the largest weights in the FCI wave
function out of a large CIPSI calculation obtained with the VDZ-BFD basis.  Within this set of determinants,
we solve the self-consistent equations of RS-DFT [see Eq.~\eqref{rs-dft-eigen-equation}]
for different values of $\mu$ using the srPBE functional. This gives the CI expansions of $\Psi^\mu$.
Then, within the same set of determinants we optimize the CI coefficients in the presence of
a simple one- and two-body Jastrow factor $e^J$ with $J = J_\text{eN} + J_\text{ee}$ and
\begin{subequations}
\begin{gather}
  J_\text{eN} = - \sum_{A=1}^{\Nat} \sum_{i=1}^{\Nelec} \qty( \frac{\alpha_A\, r_{iA}}{1 + \alpha_A\, r_{iA}} )^2,
\label{eq:jast-eN} \\
  J_\text{ee} = \sum_{i < j}^{\Nelec} \frac{a\, r_{ij}}{1 + b\, r_{ij}}. 
\label{eq:jast-ee} 
\end{gather}
\end{subequations}
The one-body Jastrow factor $J_\text{eN}$ contains the electron-nucleus terms (where $\Nat$ is the number of nuclei) with a single parameter
$\alpha_A$ per nucleus. 
The two-body Jastrow factor $J_\text{ee}$ gathers the electron-electron terms
where the sum over $i < j$ loops over all unique electron pairs. 
In Eqs.~\eqref{eq:jast-eN} and \eqref{eq:jast-ee}, $r_{iA}$ is the distance between the $i$th electron and the $A$th nucleus while $r_{ij}$ is the interlectronic distance between electrons $i$ and $j$.
The parameters $a=1/2$
and $b=0.89$ were fixed, and the parameters $\gamma_{\text{O}}=1.15$ and $\gamma_{\text{H}}=0.35$
were obtained by energy minimization of a single determinant.
The optimal CI expansion $\Psi^J$ is obtained by sampling the matrix elements
of the Hamiltonian ($\mathbf{H}$) and overlap ($\mathbf{S}$) matrices in the
basis of Jastrow-correlated determinants $e^J D_i$:
\begin{subequations}
\begin{gather}
H_{ij} = \expval{ \frac{e^J D_i}{\Psi^J}\, \frac{\hat{H}\, (e^J D_j)}{\Psi^J}  },
\\ 
S_{ij} = \expval{ \frac{e^J D_i}{\Psi^J}\, \frac{e^J D_j}{\Psi^J}  },
\end{gather}
\end{subequations}
and solving Eq.~\eqref{eq:ci-j}.\cite{Nightingale_2001}

\begin{figure*}
  \includegraphics[width=\columnwidth]{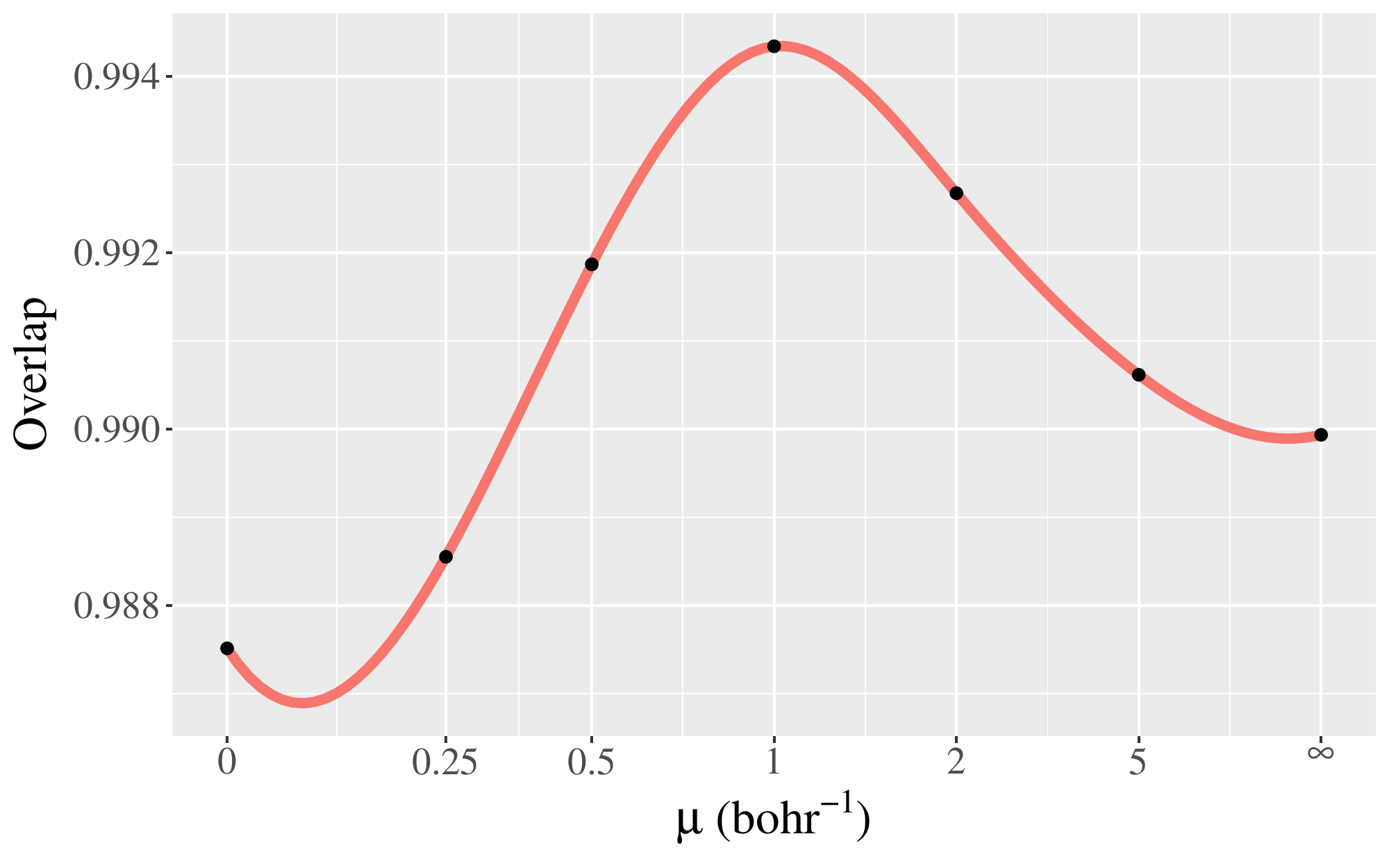}
  \includegraphics[width=\columnwidth]{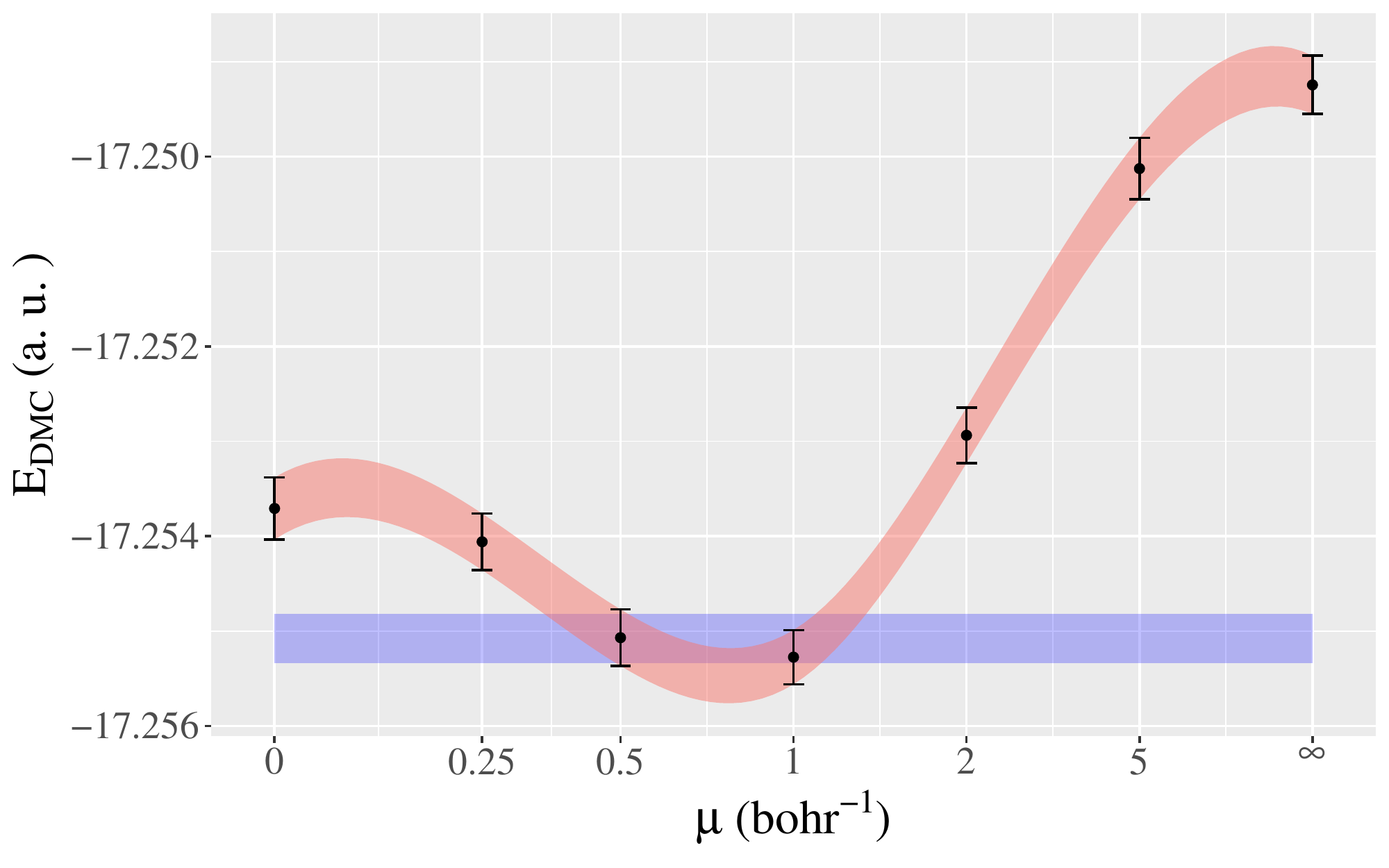}
  \caption{Left: Overlap between $\Psi^\mu$ and $\Psi^J$ as a function of $\mu$ for \ce{H2O}.
    Right: FN-DMC energy of $\Psi^\mu$ (red curve) as a function of $\mu$, together with
    the FN-DMC energy of $\Psi^J$ (blue line) for \ce{H2O}. 
    The width of the lines represent the statistical error bars.
    For these two trial wave functions, the CI expansion consists of the 200 most important
    determinants of the FCI expansion obtained with the VDZ-BFD basis (see Sec.~\ref{sec:rsdft-j} for more details). The raw data can be found in the {\SI}.}
  \label{fig:overlap}
\end{figure*}

We can easily compare $\Psi^\mu$ and  $\Psi^J$ as they are developed
on the same set of Slater determinants.
In Fig.~\ref{fig:overlap}, we plot the overlap
$\braket*{\Psi^J}{\Psi^\mu}$ obtained for water as a function of $\mu$ (left graph)
as well as the FN-DMC energy of the wave function
$\Psi^\mu$ as a function of $\mu$ together with that of $\Psi^J$ (right graph).

As evidenced by Fig.~\ref{fig:overlap}, there is a clear maximum overlap between the two trial wave functions at $\mu=1$~bohr$^{-1}$, which
coincides with the minimum of the FN-DMC energy of $\Psi^\mu$.
Also, it is interesting to notice that the FN-DMC energy of $\Psi^J$ is compatible
with that of $\Psi^\mu$ for $0.5 < \mu < 1$~bohr$^{-1}$, as shown by the overlap between the red and blue bands.
This confirms that introducing short-range correlation with DFT has
an impact on the CI coefficients similar to a Jastrow factor.
This is another key result of the present study.

\begin{figure*}
  \includegraphics[width=\columnwidth]{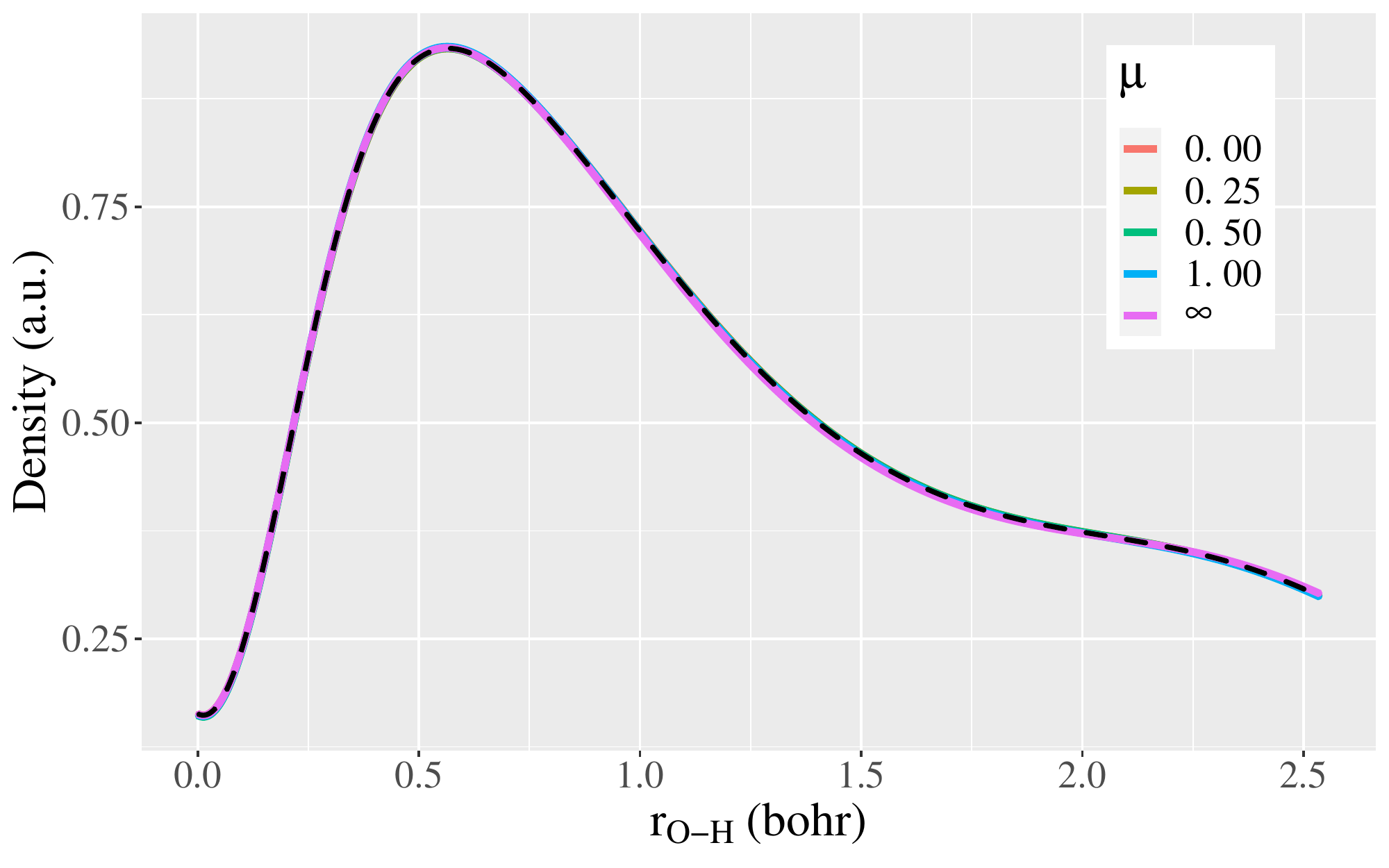}
  \includegraphics[width=\columnwidth]{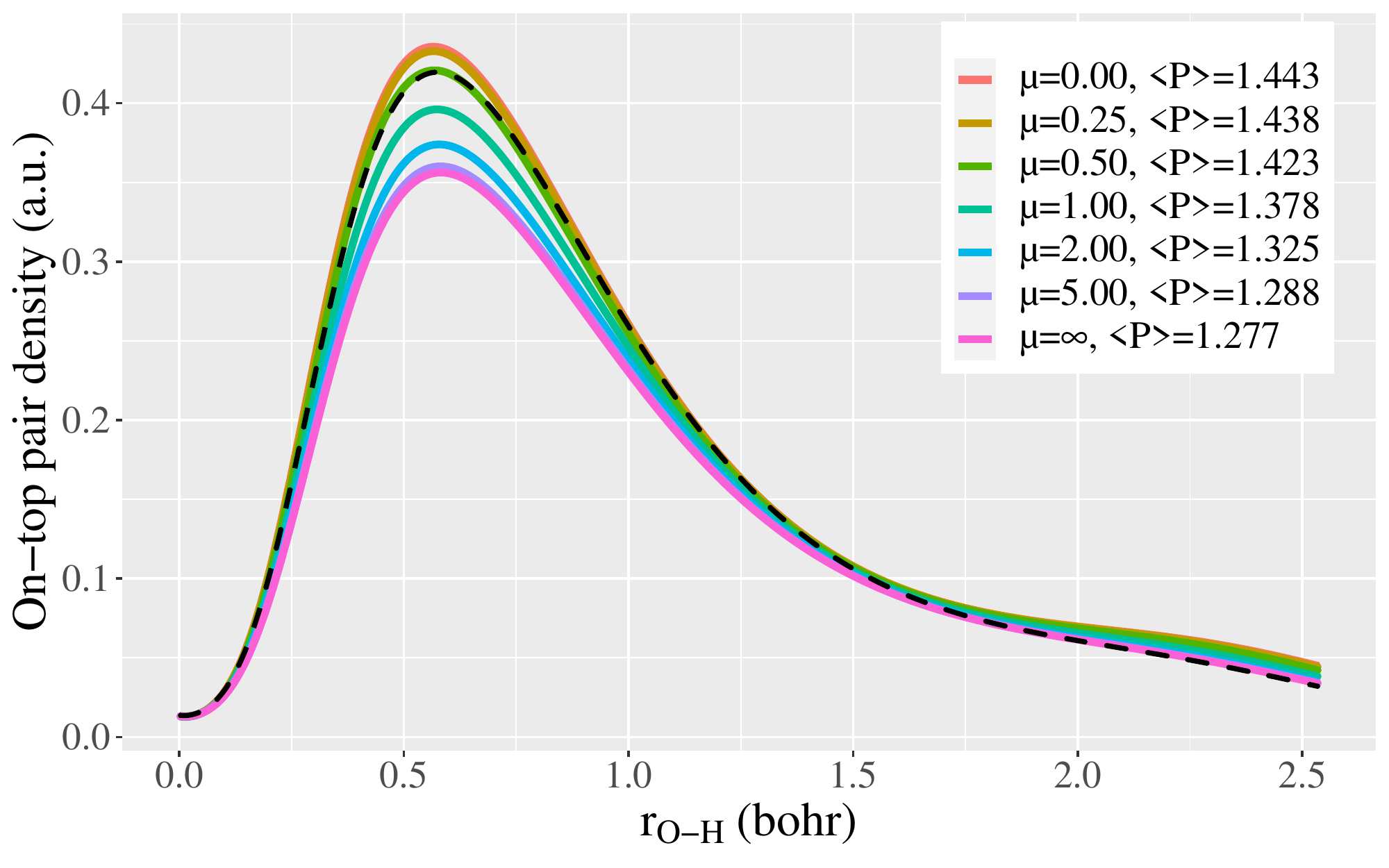}
  \caption{One-electron density $n(\br)$ (left) and on-top pair
    density $n_2(\br,\br)$ (right) along the \ce{O-H} axis of \ce{H2O}
    as a function of $\mu$ for $\Psi^\mu$, and $\Psi^J$ (dashed
    curve).
    The integrated on-top pair density $\expval{P}$ is
    given in the legend.
    For all trial wave functions, the CI expansion consists of the 200 most important
    determinants of the FCI expansion obtained with the VDZ-BFD basis (see Sec.~\ref{sec:rsdft-j} for more details). The raw data can be found in the {\SI}.}
  \label{fig:densities}
\end{figure*}

In order to refine the comparison between $\Psi^\mu$ and $\Psi^J$, we
report several quantities related to the one- and two-body densities of
$\Psi^J$ and $\Psi^\mu$ with different values of $\mu$.  First, we
report in the legend of the right panel of Fig~\ref{fig:densities} the integrated on-top pair density
\begin{equation}
 \expval{ P } = \int d\br \,n_2(\br,\br),
\end{equation}
obtained for both $\Psi^\mu$ and $\Psi^J$,
where $n_2(\br_1,\br_2)$ is the two-body density [normalized to $\Nelec(\Nelec-1)$].
Then, in order to have a pictorial representation of both the one-body density $n(\br)$ and the on-top
pair density $n_2(\br,\br)$, we report in Fig.~\ref{fig:densities}
the plots of $n(\br)$ and $n_2(\br,\br)$ along one of the \ce{O-H} axis of the water molecule.

From these data, one can clearly notice several trends.
First, the integrated on-top pair density $\expval{ P }$ decreases when $\mu$ increases,
which is expected as the two-electron interaction increases in
$H^\mu[n]$.
Second, Fig.~\ref{fig:densities} shows that the relative variations of the on-top pair density with respect to $\mu$
are much more important than that of the one-body density, the latter
being essentially unchanged between $\mu=0$ and $\mu=\infty$ while the
former can vary by about 10$\%$ in some regions. 
In the high-density region of the \ce{O-H} bond, the value of the on-top
pair density obtained from $\Psi^J$ is superimposed with
$\Psi^{\mu=0.5}$, and at a large distance the on-top pair density of $\Psi^J$ is
the closest to that of $\Psi^{\mu=\infty}$. The integrated on-top pair density
obtained with $\Psi^J$ is $\expval{P}=1.404$, which nestles between the values obtained at
$\mu=0.5$ and $\mu=1$~bohr$^{-1}$, consistently with the FN-DMC energies
and the overlap curve depicted in Fig.~\ref{fig:overlap}.

These data suggest that the wave functions $\Psi^{0.5 \le \mu \le 1}$ and $\Psi^J$ are close,
and therefore that the operators that produced these wave functions (\ie, $H^\mu[n]$ and $e^{-J}He^J$) contain similar physics.
Considering the form of $\hat{H}^\mu[n]$ [see Eq.~\eqref{H_mu}],
one can notice that the differences with respect to the usual bare Hamiltonian come
from the non-divergent two-body interaction $\hat{W}_{\text{ee}}^{\text{lr},\mu}$
and the effective one-body potential $\hat{\bar{V}}_{\text{Hxc}}^{\text{sr},\mu}[n]$ which is the functional derivative of the Hxc functional.
The roles of these two terms are therefore very different: with respect
to the exact ground-state wave function $\Psi$, the non-divergent two-body interaction
increases the probability of finding electrons at short distances in $\Psi^\mu$,
while the effective one-body potential $\hat{\bar{V}}_{\text{Hxc}}^{\text{sr},\mu}[n_{\Psi^{\mu}}]$,
providing that it is exact, maintains the exact one-body density.
This is clearly what has been observed in
Fig.~\ref{fig:densities}.
Regarding now the transcorrelated Hamiltonian $e^{-J}He^J$, as pointed out by Ten-no,\cite{Tenno_2000}
the effective two-body interaction induced by the presence of a Jastrow factor
can be non-divergent when a proper two-body Jastrow factor $J_\text{ee}$ is chosen, \ie, the Jastrow factor must fulfill the so-called electron-electron cusp conditions. \cite{Kato_1957,Pack_1966}
There is therefore a clear parallel between $\hat{W}_{\text{ee}}^{\text{lr},\mu}$ in RS-DFT and $J_\text{ee}$ in FN-DMC.
Moreover, the one-body Jastrow term $J_\text{eN}$ ensures that the one-body density remains unchanged when the CI coefficients are re-optmized in the presence of $J_\text{ee}$.
There is then a second clear parallel between $\hat{\bar{V}}_{\text{Hxc}}^{\text{sr},\mu}[n]$ in RS-DFT and $J_\text{eN}$ in FN-DMC.
Thus, one can understand the similarity between the eigenfunctions of $H^\mu$ and the optimization of the Slater-Jastrow wave function:
they both deal with an effective non-divergent interaction but still
produce a reasonable one-body density.

\subsection{Intermediate conclusions}
\label{sec:int_ccl}

As conclusions of the first part of this study, we can highlight the following observations:
\begin{itemize}
\item With respect to the nodes of a KS determinant or a FCI wave function,
  one can obtain a multi-determinant trial wave function $\Psi^\mu$ with a smaller
  fixed-node error by properly choosing an optimal value of $\mu$.
\item The optimal $\mu$ value is system- and basis-set-dependent, and it grows with basis set size.
\item Numerical experiments (overlap $\braket*{\Psi^\mu}{\Psi^J}$, 
  one-body density, on-top pair density, and FN-DMC energy) indicate
  that the RS-DFT scheme essentially plays the role of a simple Jastrow factor
  by mimicking short-range correlation effects.  This latter
  statement can be qualitatively understood by noticing that both RS-DFT
  and the trans-correlated approach deal with an effective non-divergent
  electron-electron interaction, while keeping the density constant.
\end{itemize}

\section{Energy differences in FN-DMC: atomization energies}
\label{sec:atomization}

Atomization energies are challenging for post-HF methods
because their calculation requires a subtle balance in the
description of atoms and molecules. The mainstream one-electron basis sets employed in molecular electronic structure 
calculations are atom-centered, so they are, by construction, better adapted to
atoms than molecules. Thus, atomization energies usually tend to be
underestimated by variational methods.
In the context of FN-DMC calculations, the nodal surface is imposed by
the determinantal part of the trial wave function which is expanded in the very same atom-centered basis
set. Thus, we expect the fixed-node error to be also intimately connected to
the basis set incompleteness error.
Increasing the size of the basis set improves the description of
the density and of the electron correlation, but also reduces the
imbalance in the description of atoms and
molecules, leading to more accurate atomization energies.
The size-consistency and the spin-invariance of the present scheme, 
two key properties to obtain accurate atomization energies, 
are discussed in Appendices \ref{app:size} and \ref{app:spin}, respectively.

\begin{squeezetable}
\begin{table*}
  \caption{Mean absolute errors (MAEs), mean signed errors (MSEs), and
    root mean square errors (RMSEs) with respect to the NIST reference values obtained with various methods and
    basis sets.
    All quantities are given in kcal/mol. The raw data can be found in the {\SI}.}
  \label{tab:mad}
  \begin{ruledtabular}
    \begin{tabular}{ll ddd ddd ddd}
            &       & \mc{3}{c}{VDZ-BFD}                &  \mc{3}{c}{VTZ-BFD}    & \mc{3}{c}{VQZ-BFD}   \\
            \cline{3-5} \cline{6-8} \cline{9-11}
Method        &  $\mu$  &  \tabc{MAE}                   &  \tabc{MSE}                    &  \tabc{RMSE}                  &  \tabc{MAE}         &  \tabc{MSE}        &  \tabc{RMSE}        &  \tabc{MAE}         &  \tabc{MSE}        &  \tabc{RMSE}        \\
\hline
PBE               &  0           &  5.02                  &  -3.70                  &  6.04                  &  4.57        &  1.00       &  5.32        &  5.31        &  0.79       &  6.27        \\
BLYP              &  0           &  9.53                  &  -9.21                  &  7.91                  &  5.58        &  -4.44      &  5.80        &  5.86        &  -4.47      &  6.43        \\
PBE0              &  0           &  11.20                 &  -10.98                 &  8.68                  &  6.40        &  -5.78      &  5.49        &  6.28        &  -5.65      &  5.08        \\
B3LYP             &  0           &  11.27                 &  -10.98                 &  9.59                  &  7.27        &  -5.77      &  6.63        &  6.75        &  -5.53      &  6.09        \\
\\
CCSD(T)           &  \(\infty\)  &  24.10                 &  -23.96                 &  13.03                 &  9.11        &  -9.10      &  5.55        &  4.52        &  -4.38      &  3.60        \\
\\
RS-DFT-CIPSI      &  0           &  4.53                  &  -1.66                  &  5.91                  &  6.31        &  0.91       &  7.93        &  6.35        &  3.88       &  7.20        \\
           &  1/4         &  5.55                  &  -4.66                  &  5.52                  &  4.58        &  1.06       &  5.72        &  5.48        &  1.52       &  6.93        \\
           &  1/2         &  13.42                 &  -13.27                 &  7.36                  &  6.77        &  -6.71      &  4.56        &  6.35        &  -5.89      &  5.18        \\
           &  1           &  17.07                 &  -16.92                 &  9.83                  &  9.06        &  -9.06      &  5.88        &           &          &           \\
           &  2           &  19.20                 &  -19.05                 &  10.91                 &           &          &           &           &          &           \\
           &  5           &  22.93                 &  -22.79                 &  13.24                 &           &          &           &           &          &           \\
           &  \(\infty\)  &  23.63(4)              &  -23.49(4)              &  12.81(4)              &  8.43(39)    &  -8.43(39)  &  4.87(7)     &  4.51(78)    &  -4.18(78)  &  4.19(20)         \\
\\
DMC@  &  0           &  4.61(34)  &  -3.62(34)  &  5.30(09)  &  3.52(19)    &  -1.03(19)  &  4.39(04)    &  3.16(26)    &  -0.12(26)  &  4.12(03)   \\
 RS-DFT-CIPSI                 &  1/4         &  4.04(37)  &  -3.13(37)  &  4.88(10)  &  3.39(77)    &  -0.59(77)  &  4.44(34)    &  2.90(25)    &  0.25(25)   &  3.745(5)   \\
                  &  1/2         &  3.74(35)  &  -3.53(35)  &  4.03(23)  &  2.46(18)    &  -1.72(18)  &  3.02(06)    &  2.06(35)    &  -0.44(35)  &  2.74(13)    \\
                  &  1           &  5.42(29)  &  -5.14(29)  &  4.55(03)  &  4.38(94)    &  -4.24(94)  &  5.11(31)    &           &          &           \\
                  &  2           &  5.98(83)  &  -5.91(83)  &  4.79(71)  &              &         &           &           &          &           \\
                  &  5           &  6.18(84)  &  -6.13(84)  &  4.87(55)  &             &          &           &           &          &           \\
                  &  \(\infty\)  &  7.38(1.08)            &  -7.38(1.08)               &  5.67(68)  &           &          &           &           &          &           \\
                  &  Opt.        &  5.85(1.75)            &  -5.63(1.75)            &  4.79(1.11)            &           &          &           &           &          &           \\
    \end{tabular}
  \end{ruledtabular}
\end{table*}
\end{squeezetable}

The atomization energies of the 55 molecules of the Gaussian-1
theory\cite{Pople_1989,Curtiss_1990} were chosen as a benchmark set to test the
performance of the RS-DFT-CIPSI trial wave functions in the context of
energy differences. 
Calculations were made in the double-, triple-
and quadruple-$\zeta$ basis sets with different values of $\mu$, and using
NOs from a preliminary CIPSI calculation as a starting point (see Fig.~\ref{fig:algo}).
\footnote{At $\mu=0$, the number of determinants is not equal to one because
we have used the natural orbitals of a preliminary CIPSI calculation, and
not the srPBE orbitals.
So the Kohn-Sham determinant is expressed as a linear combination of
determinants built with NOs. It is possible to add
an extra step to the algorithm to compute the NOs from the
RS-DFT-CIPSI wave function, and re-do the RS-DFT-CIPSI calculation with
these orbitals to get an even more compact expansion. In that case, we would
have converged to the KS orbitals with $\mu=0$, and the
solution would have been the PBE single determinant.}
For comparison, we have computed the energies of all the atoms and
molecules at the KS-DFT level with various semi-local and hybrid density functionals [PBE, \cite{PerBurErn-PRL-96} BLYP, \cite{Becke_1988,Lee_1988} PBE0, \cite{Perdew_1996} and B3LYP \cite{Becke_1993}], and at
the CCSD(T) level. \cite{Cizek_1969,Purvis_1982,Scuseria_1988,Scuseria_1989} Table~\ref{tab:mad} gives the corresponding mean
absolute errors (MAEs), mean signed errors (MSEs), and root mean square errors (RMSEs) 
with respect to the NIST reference values as explained in Sec.~\ref{sec:comp-details}.
For FCI (RS-DFT-CIPSI, $\mu=\infty$) we have
provided the extrapolated values (\ie, when $\EPT \to 0$), and, although one cannot provide theoretically sound error bars, they
correspond here to the difference between the extrapolated energies computed with a
two-point and a three-point linear extrapolation. \cite{Loos_2018a,Loos_2019,Loos_2020b,Loos_2020c}

In this benchmark, the great majority of the systems are weakly correlated and are then well
described by a single determinant. Therefore, the atomization energies
calculated at the KS-DFT level are relatively accurate, even when
the basis set is small. The introduction of exact exchange (B3LYP and
PBE0) makes the results more sensitive to the basis set, and reduce the
accuracy.  Note that, due to the approximate nature of the xc functionals, 
the statistical quantities associated with KS-DFT atomization energies do not converge towards zero and remain altered even in the CBS limit.
Thanks to the single-reference character of these systems,
the CCSD(T) energy is an excellent estimate of the FCI energy, as
shown by the very good agreement of the MAE, MSE and RMSE of CCSD(T)
and FCI energies for each basis set.
The imbalance in the description of molecules compared
to atoms is exhibited by a very negative value of the MSE for
CCSD(T)/VDZ-BFD ($-23.96$ kcal/mol) and FCI/VDZ-BFD ($-23.49\pm0.04$ kcal/mol), which is reduced by a factor of two
when going to the triple-$\zeta$ basis, and again by a factor of two when
going to the quadruple-$\zeta$ basis.

This significant imbalance at the VDZ-BFD level affects the nodal
surfaces, because although the FN-DMC energies obtained with near-FCI
trial wave functions are much lower than the FN-DMC
energies at $\mu = 0$, the MAE obtained with FCI ($7.38\pm1.08$ kcal/mol) is
larger than the MAE at $\mu = 0$ ($4.61\pm0.34$ kcal/mol).
Using the FCI trial wave function the MSE is equal to the
negative MAE which confirms that the atomization energies are systematically
underestimated. This corroborates that some of the basis set
incompleteness error is transferred in the fixed-node error.

Within the double-$\zeta$ basis set, the calculations could be performed for the
whole range of values of $\mu$, and the optimal value of $\mu$ for the
trial wave function was estimated for each system by searching for the
minimum of the spline interpolation curve of the FN-DMC energy as a
function of $\mu$.
This corresponds to the line labelled as ``Opt.'' in Table~\ref{tab:mad}.
The optimal $\mu$ value for each system is reported in the \SI.
Using the optimal value of $\mu$ clearly improves the
MAEs, MSEs, and RMSEs as compared to the FCI wave function. This
result is in line with the common knowledge that re-optimizing
the determinantal component of the trial wave function in the presence
of electron correlation reduces the errors due to the basis set incompleteness.
These calculations were done only for the smallest basis set
because of the expensive computational cost of the QMC calculations
when the trial wave function contains more than a few million
determinants. \cite{Scemama_2016}
At the RS-DFT-CIPSI/VTZ-BFD level, one can see that 
the MAEs are larger for $\mu=1$~bohr$^{-1}$ ($9.06$ kcal/mol) than for
FCI ($8.43\pm0.39$ kcal/mol).
The same comment applies to $\mu=0.5$~bohr$^{-1}$ with the quadruple-$\zeta$ basis.

\begin{figure*}
  \centering
  \includegraphics[width=\textwidth]{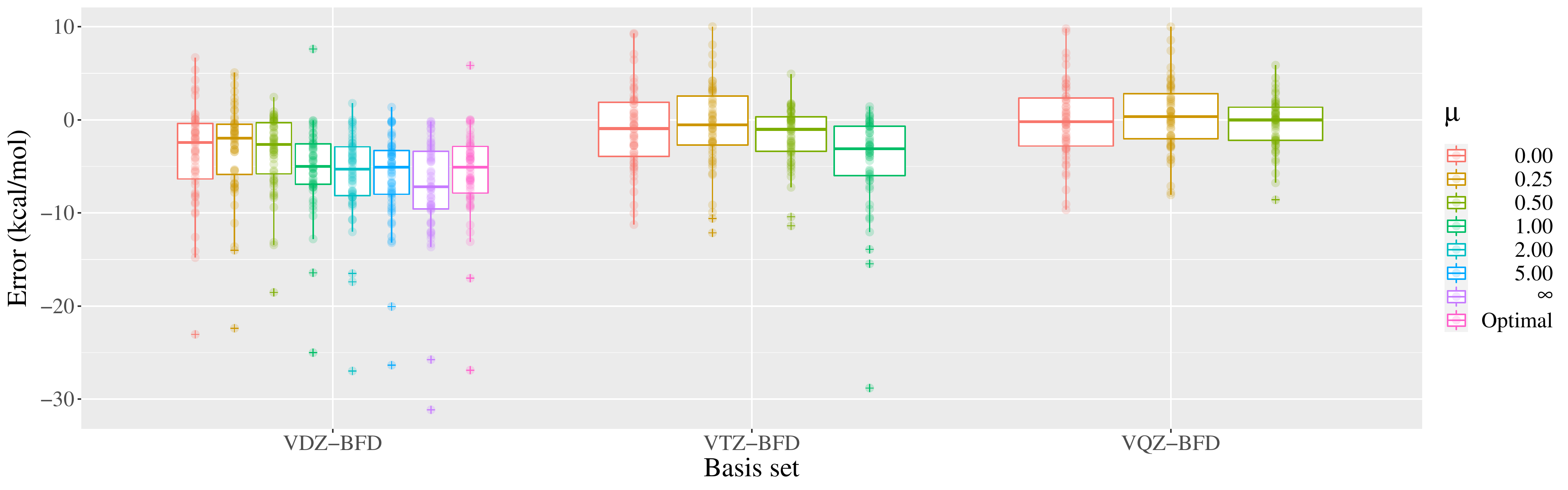}
  \caption{Errors in the FN-DMC atomization energies (in kcal/mol) for various
    trial wave functions $\Psi^{\mu}$ and basis sets. Each dot corresponds to an atomization
    energy.
    The boxes contain the data between first and third quartiles, and
    the line in the box represents the median. The outliers are shown
    with a cross. The raw data can be found in the {\SI}.}
  \label{fig:g2-dmc}
\end{figure*}

Searching for the optimal value of $\mu$ may be too costly and time consuming, so we have
computed the MAEs, MSEs and RMSEs for fixed values of $\mu$. 
As illustrated in Fig.~\ref{fig:g2-dmc} and Table \ref{tab:mad}, 
the best choice for a fixed value of $\mu$ is
$0.5$ bohr$^{-1}$ for all three basis sets. It is the value for which
the MAE [$3.74(35)$, $2.46(18)$, and $2.06(35)$ kcal/mol] and RMSE [$4.03(23)$,
$3.02(06)$, and $2.74(13)$ kcal/mol] are minimal. Note that these values
are even lower than those obtained with the optimal value of
$\mu$. Although the FN-DMC energies are higher, the numbers show that
they are more consistent from one system to another, giving improved
cancellations of errors.
This is yet another key result of the present study, and it can be explained by the lack of size-consistency when one considers different $\mu$ values for the molecule and the isolated atoms.
This observation was also mentioned in the context of optimally-tune range-separated hybrids. \cite{Stein_2009,Karolewski_2013,Kronik_2012}

\begin{figure*}
  \centering
  \includegraphics[width=\textwidth]{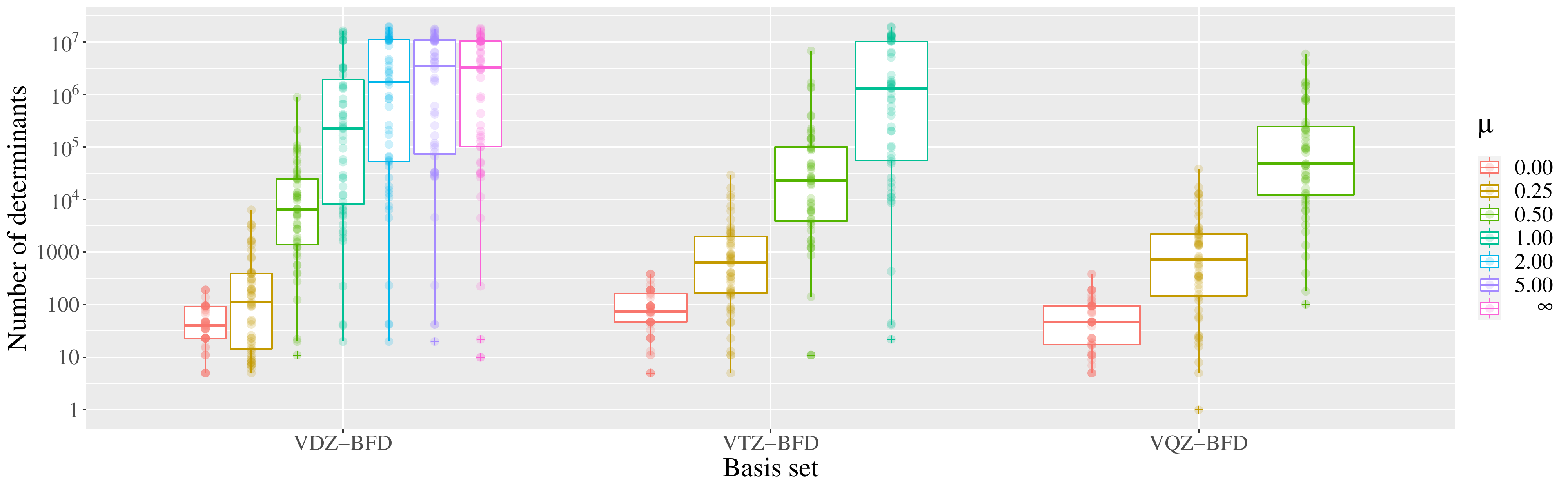}
  \caption{Number of determinants for various trial wave
    functions $\Psi^{\mu}$ and basis sets. Each dot corresponds to an atomization energy.
    The boxes contain the data between first and third quartiles, and
    the line in the box represents the median. The outliers are shown
    with a cross. The raw data can be found in the {\SI}.}
  \label{fig:g2-ndet}
\end{figure*}

The number of determinants in the trial wave functions are shown in
Fig.~\ref{fig:g2-ndet}. As expected, the number of determinants
is smaller when $\mu$ is small and larger when $\mu$ is large.
It is important to note that the median of the number of
determinants when $\mu=0.5$~bohr$^{-1}$ is below $100\,000$ determinants
with the VQZ-BFD basis, making these calculations feasible
with such a large basis set. At the double-$\zeta$ level, compared to the
FCI trial wave functions, the median of the number of determinants is
reduced by more than two orders of magnitude.
Moreover, going to $\mu=0.25$~bohr$^{-1}$ gives a median close to 100
determinants at the VDZ-BFD level, and close to $1\,000$ determinants
at the quadruple-$\zeta$ level for only a slight increase of the
MAE. Hence, RS-DFT-CIPSI trial wave functions with small values of
$\mu$ could be very useful for large systems to go beyond the
single-determinant approximation at a very low computational cost
while ensuring size-consistency.
For the largest systems, as shown in Fig.~\ref{fig:g2-ndet},
there are many systems for which we could not reach the threshold
$\EPT<1$~m\hartree{} as the number of determinants exceeded
10~million before this threshold was reached.
For these cases, there is then a
small size-consistency error originating from the imbalanced
truncation of the wave functions, which is not present in the
extrapolated FCI energies (see Appendix \ref{app:size}). 

\section{Conclusion}
\label{sec:conclusion}

In the present work, we have shown that introducing short-range correlation via
a range-separated Hamiltonian in a FCI expansion yields improved
nodal surfaces, especially with small basis sets. The effect of short-range DFT
on the determinant expansion is similar to the effect of re-optimizing
the CI coefficients in the presence of a Jastrow factor, but without
the burden of performing a stochastic optimization.

In addition to the intermediate conclusions drawn in Sec.~\ref{sec:int_ccl}, 
we have shown that varying the range-separation parameter $\mu$ and approaching
RS-DFT-FCI with CIPSI provides a way to adapt the number of
determinants in the trial wave function, leading to
size-consistent FN-DMC energies.
We propose two methods. The first one is for the computation of
accurate total energies by a one-parameter optimization of the FN-DMC
energy via the variation of the parameter $\mu$.
The second method is for the computation of energy differences, where
the target is not the lowest possible FN-DMC energies but the best
possible cancellation of errors. Using a fixed value of $\mu$
increases the (size-)consistency of the trial wave functions, and we have found
that $\mu=0.5$~bohr$^{-1}$ is the value where the cancellation of
errors is the most effective.
Moreover, such a small value of $\mu$ gives extremely
compact wave functions, making this recipe a good candidate for
the accurate description of the whole potential energy surfaces of
large systems. If the number of determinants is still too large, the
value of $\mu$ can be further reduced to $0.25$~bohr$^{-1}$ to get
extremely compact wave functions at the price of less efficient
cancellations of errors.

We hope to report, in the near future, a detailed investigation of strongly-correlated systems with the present RS-DFT-CIPSI scheme.

\begin{acknowledgments}
A.B was supported by the U.S.~Department of Energy, Office of Science, Basic Energy Sciences, Materials Sciences and Engineering Division, as part of the Computational Materials Sciences Program and Center for Predictive Simulation of Functional Materials.
This work was performed using HPC resources from GENCI-TGCC (Grand
Challenge 2019-gch0418) and from CALMIP (Toulouse) under allocation
2020-18005.
Funding from \textit{``Projet International de Coop\'eration Scientifique''} (PICS08310) and from the \textit{``Centre National de la Recherche Scientifique''} is acknowledged.
This study has been (partially) supported through the EUR grant NanoX No.~ANR-17-EURE-0009 in the framework of the \textit{``Programme des Investissements d'Avenir''}.
\end{acknowledgments}

\section*{Data availability}

The data that support the findings of this study are openly available in Zenodo at \url{http://doi.org/10.5281/zenodo.3996568}.

\appendix

\section{Size consistency}
\label{app:size}

An extremely important feature required to get accurate
atomization energies is size-consistency (or strict separability),
since the numbers of correlated electron pairs in the molecule and its isolated atoms
are different.

KS-DFT energies are size-consistent, and because xc functionals are 
directly constructed in complete basis, their convergence with respect 
to the size of the basis set is relatively fast. \cite{FraMusLupTou-JCP-15,Giner_2018,Loos_2019d,Giner_2020} 
Hence, DFT methods are very well adapted to
the calculation of atomization energies, especially with small basis
sets. \cite{Giner_2018,Loos_2019d,Giner_2020} 
However, in the CBS, KS-DFT atomization energies do not match the exact values due to the approximate nature of the xc functionals.

Likewise, FCI is also size-consistent, but the convergence of
the FCI energies towards the CBS limit is much slower because of the
description of short-range electron correlation using atom-centered
functions. \cite{Kutzelnigg_1985,Kutzelnigg_1991,Hattig_2012} 
Eventually though, the exact atomization energies will be reached.

In the context of SCI calculations, when the variational energy is
extrapolated to the FCI energy \cite{Holmes_2017} there is no
size-consistency error. But when the truncated SCI wave function is used
as a reference for post-HF methods such as SCI+PT2
or for QMC calculations, there is a residual size-consistency error
originating from the truncation of the wave function.

QMC energies can be made size-consistent by extrapolating the
FN-DMC energy to estimate the energy obtained with the FCI as a trial
wave function.\cite{Scemama_2018,Scemama_2018b} Alternatively, the
size-consistency error can be reduced by choosing the number of
selected determinants such that the sum of the PT2 corrections on the
fragments is equal to the PT2 correction of the molecule, enforcing that
the variational potential energy surface (PES) is
parallel to the perturbatively corrected PES, which is a relatively
accurate estimate of the FCI PES.\cite{Giner_2015}

Another source of size-consistency error in QMC calculations originates
from the Jastrow factor. Usually, the Jastrow factor contains
one-electron, two-electron and one-nucleus-two-electron terms.
The problematic part is the two-electron term, whose simplest form can
be expressed as in Eq.~\eqref{eq:jast-ee}.
The parameter
$a$ is determined by the electron-electron cusp condition, \cite{Kato_1957,Pack_1966} and $b$ is obtained by energy
or variance minimization.\cite{Coldwell_1977,Umrigar_2005}
One can easily see that this parameterization of the two-body
interaction is not size-consistent: the dissociation of a
diatomic molecule \ce{AB} with a parameter $b_{\ce{AB}}$
will lead to two different two-body Jastrow factors, each
with its own optimal value $b_{\ce{A}}$ and $b_{\ce{B}}$. To remove the
size-consistency error on a PES using this ans\"atz for $J_\text{ee}$,
one needs to impose that the parameters of $J_\text{ee}$ are fixed, \ie,
$b_{\ce{A}} = b_{\ce{B}} = b_{\ce{AB}}$.

When pseudopotentials are used in a QMC calculation, it is of common
practice to localize the non-local part of the pseudopotential on the
complete trial wave function $\Phi$.
If the wave function is not size-consistent,
so will be the locality approximation. Within the DLA,\cite{Zen_2019} the Jastrow factor is
removed from the wave function on which the pseudopotential is localized.
The great advantage of this approximation is that the FN-DMC energy
only depends on the parameters of the determinantal component. Using a
non-size-consistent Jastrow factor, or a non-optimal Jastrow factor will
not introduce an additional error in FN-DMC calculations, although it
will reduce the statistical errors by reducing the variance of the
local energy. Moreover, the integrals involved in the pseudopotential
are computed analytically and the computational cost of the
pseudopotential is dramatically reduced (for more details, see
Ref.~\onlinecite{Scemama_2015}).

In this section, we make a numerical verification that the produced
wave functions are size-consistent for a given range-separation
parameter.
We have computed the FN-DMC energy of the dissociated fluorine dimer, where
the two atoms are separated by 50~\AA.  We expect that the energy
of this system is equal to twice the energy of the fluorine atom.
The data in Table~\ref{tab:size-cons} shows that this is indeed the
case, so we can conclude that the proposed scheme provides
size-consistent FN-DMC energies for all values of $\mu$ (within
twice the statistical error bars).

\begin{table}
  \caption{FN-DMC energy (in \hartree{}) using the VDZ-BFD basis set and the srPBE functional
    of the fluorine atom and the dissociated \ce{F2} molecule for various $\mu$ values.
    The size-consistency error is also reported.}
  \label{tab:size-cons}
  \begin{ruledtabular}
    \begin{tabular}{cccc}
      $\mu$ &  \ce{F}                &  Dissociated \ce{F2}  & Size-consistency error \\
      \hline
      0.00  &  $-24.188\,7(3)$  & $-48.377\,7(3)$     & $-0.000\,3(4)$ \\
      0.25  &  $-24.188\,7(3)$  & $-48.377\,2(4)$     & $+0.000\,2(5)$ \\
      0.50  &  $-24.188\,8(1)$  & $-48.376\,9(4)$     & $+0.000\,7(4)$ \\
      1.00  &  $-24.189\,7(1)$  & $-48.380\,2(4)$     & $-0.000\,8(4)$ \\
      2.00  &  $-24.194\,1(3)$  & $-48.388\,4(4)$     & $-0.000\,2(5)$ \\
      5.00  &  $-24.194\,7(4)$  & $-48.388\,5(7)$     & $+0.000\,9(8)$ \\
   $\infty$ &  $-24.193\,5(2)$  & $-48.386\,9(4)$     & $+0.000\,1(5)$ \\
    \end{tabular}
  \end{ruledtabular}
\end{table}

\section{Spin invariance}
\label{app:spin}

Closed-shell molecules often dissociate into open-shell
fragments. To get reliable atomization energies, it is important to
have a theory which is of comparable quality for open- and
closed-shell systems. A good check is to make sure that all the components
of a spin multiplet are degenerate, as expected from exact solutions.

FCI wave functions have this property and yield degenerate energies with
respect to the spin quantum number $m_s$. 
However, multiplying the determinantal part of the trial wave function by a
Jastrow factor introduces spin contamination if the Jastrow parameters
for the same-spin electron pairs are different from those
for the opposite-spin pairs.\cite{Tenno_2004}
Again, when pseudopotentials are employed, this tiny error is transferred
to the FN-DMC energy unless the DLA is enforced.

The context is rather different within KS-DFT.
Indeed, mainstream density functionals have distinct functional forms to take 
into account correlation effects of same-spin and opposite-spin electron pairs. 
Therefore, KS determinants corresponding to different values of $m_s$ lead to different total energies.
Consequently, in the context of RS-DFT, the determinant expansion is impacted by this spurious effect, as opposed to FCI.

In this Appendix, we investigate the impact of the spin contamination on the FN-DMC energy
originating from the short-range density functional. We have
computed the energies of the carbon atom in its triplet state
with the VDZ-BFD basis set and the srPBE functional. 
The calculations are performed for $m_s=1$ (3 spin-up
and 1 spin-down electrons) and for $m_s=0$ (2 spin-up and 2
spin-down electrons).

The results are reported in Table~\ref{tab:spin}.
Although the energy obtained with $m_s=0$ is higher than the one obtained with $m_s=1$, the
bias is relatively small, \ie, more than one order of magnitude smaller
than the energy gained by reducing the fixed-node error going from the single
determinant to the FCI trial wave function. The largest spin-invariance error, close to
$2$ m\hartree{}, is obtained for $\mu=0$, but this bias decreases quickly
below $1$ m\hartree{} when $\mu$ increases. As expected, with $\mu=\infty$
we observe a perfect spin-invariance of the energy (within the error bars), and the bias is not
noticeable for $\mu=5$~bohr$^{-1}$.

Hence, at the FN-DMC level, the error due to
the spin invariance with RS-DFT-CIPSI trial wave functions is below the
chemical accuracy threshold, and is not expected to be problematic for the
comparison of atomization energies.

\begin{table}
  \caption{FN-DMC energy (in \hartree{}) for various $\mu$ values of the triplet carbon atom with
    different values of $m_s$ computed with the VDZ-BFD basis set and the srPBE functional.    
    The spin-invariance error is also reported.}
  \label{tab:spin}
  \begin{ruledtabular}
    \begin{tabular}{cccc}
      $\mu$ &  $m_s=1$          & $m_s=0$             & Spin-invariance error \\
      \hline
      0.00  &  $-5.416\,8(1)$  & $-5.414\,9(1)$     & $+0.001\,9(2)$ \\
      0.25  &  $-5.417\,2(1)$  & $-5.416\,5(1)$     & $+0.000\,7(1)$ \\
      0.50  &  $-5.422\,3(1)$  & $-5.421\,4(1)$     & $+0.000\,9(2)$ \\
      1.00  &  $-5.429\,7(1)$  & $-5.429\,2(1)$     & $+0.000\,5(2)$ \\
      2.00  &  $-5.432\,1(1)$  & $-5.431\,4(1)$     & $+0.000\,7(2)$ \\
      5.00  &  $-5.431\,7(1)$  & $-5.431\,4(1)$     & $+0.000\,3(2)$ \\
   $\infty$ &  $-5.431\,6(1)$  & $-5.431\,3(1)$     & $+0.000\,3(2)$ \\
    \end{tabular}
  \end{ruledtabular}
\end{table}

\bibliography{rsdft-cipsi-qmc}

\end{document}